\documentclass{elsart41}
\usepackage{natbib}
\usepackage{graphics}
\usepackage{graphicx}
\usepackage{amssymb}
\newcommand{\ce}{CeCu$_2$Si$_2$}

\newcommand{\cps}{CePd$_2$Si$_2$}
\newcommand{\tc}{$T_c$}

\newcommand{\pv}{$P_{\rm v}$}
\bibpunct{[}{]}{,}{a}{}{;}
\begin{document}

\begin{frontmatter}



\title{Valence fluctuation mediated superconductivity in CeCu$_2$Si$_2$}


\author[JP1,CH]{Alexander T.~Holmes\corauthref{Holmes}}
\ead{alex@djebel.mp.es.osaka-u.ac.jp}
\author[CH]{Didier Jaccard}

\address[JP1]{KYOKUGEN, Osaka University, Toyonaka, Osaka
560-8531, Japan}
\address[CH]{DPMC, Section de Physique, University of Geneva, CH-1211 Gen\`{e}ve 4, Switzerland}


\corauth[Holmes]{Corresponding author. Tel: +81 (0)6 68506479 fax:
 +81 (0)6 68506662}

\begin{abstract}
It has been proposed that there are two types of superconductivity
in \ce, mediated by spin fluctuations at ambient pressure, and by
critical valence fluctuations around a charge instability at a
pressure $P_{\rm v}\simeq4.5\:$GPa. We present in detail some of
the unusual features of this novel type of superconducting state,
including the coexistence of superconductivity and huge residual
resistivity of the order of the Ioffe-Regel limit, large and
pressure dependent resistive transition widths in a single crystal
measured under hydrostatic conditions, asymmetric pressure
dependence of the specific heat jump shape, unrelated to the
resistivity width, and negative temperature dependence of the
normal state resistivity below 10$\:$K at very high pressure.

\end{abstract}

\end{frontmatter}

\section*{Introduction}

The heavy fermion (HF) family has been a constant source of
scientific interest for many years, owing to the rich variety of
electronic properties and competing ground states. These can be be
tuned within an experimentally accessible region of pressure,
magnetic field, or chemical substitution. Many of the most
interesting features of these compounds are found where the
localised $f$ electrons are on the border of itinerancy, and
strong interactions with the more loosely bound conduction
electrons have a hugely important effect.

Superconductivity is one of the most striking properties of many
HF Ce compounds.  The superconducting pairing mechanism is still
not fully understood, but the presence of a local magnetic moment
at each Ce site strongly disfavours a phonon-mediated interaction.

The leading candidate for the pairing mechanism in HF
superconductors has been mediation by low energy spin fluctuations
around a so-called magnetic quantum critical point (QCP). This is
where the magnetic ordering temperature is driven to zero at a
pressure $P_c$, if pressure is the control parameter, and quantum
fluctuations between competing ground states dominate the
properties of the system. For example in \cps\ and CeIn$_3$,
superconductivity was found in a small dome around the QCP, but
only in samples of very high purity, where the electronic mean
free path exceeded the superconducting coherence length $\xi$
\cite{Mathur98}. The nature of singularity at $P_c$ has usually
been treated as second order, however this view may not be
entirely correct \cite{Flouquet05}.

The heavy fermion superconductor CeCu$_2$Si$_2$ has a very
unusually shaped superconducting region in the
pressure-temperature phase diagram, and its isoelectronic parent
compound CeCu$_2$Ge$_2$ displays very similar behaviour under
larger compression. It was recently shown that the addition of Ge
impurities to CeCu$_2$(Si$_x$Ge$_{1-x}$)$_2$ separates the
superconducting region into two separate domes, centred on the
antiferromagnetic QCP and a valence transition respectively
\cite{Yuan03b}. We have argued that \ce\ has a second mechanism of
superconductivity at high pressure, where critical valence
fluctuations, between Ce $4f^n$ and $4\textrm{f}^{n-1}+[5\rm d6\rm
s]$ electronic configurations form the basis of the pairing
mechanism\cite{Onishi00,Holmes04a}. These are associated with a
nearly first order valence instability, analogous to the Ce
$\alpha-\gamma$ transition with a critical end point close to zero
temperature. There is a good deal of evidence for an abrupt change
in Ce valence, along with several anomalies in the normal state,
close the pressure at which \tc\ attains its maximum value of
around 2.5$\:$K.

The superconducting transition temperature also displays a very
anisotropic response to uniaxial stress\cite{Holmesunpublished},
similar to the effect seen in \cps\cite{Demuer02}. This is
reminiscent of the strong sensitivity of \tc\ to the ratio of
tetragonal lattice parameters $c/a$ seen in the Ce$M$In$_5$ ($M$
-- Co, Rh, Ir) and Pu$M$Ga$_5$ ($M$ -- Co, Rh) compounds
\cite{bauer:147005}.

If we accept that superconductivity at high pressure in \ce\ is
mediated by valence fluctuations, does it have any uniquely
identifying features? More specifically, is there any way to
recognize such a compound without an exhaustive pressure
investigation such as reported in \cite{Holmes04a}? Since \ce\ is
so far the only firm example of this phenomenon, it is hard to say
what is a characteristic feature of valence fluctuation (VF)
mediated superconductivity, and what is unique to \ce. We have
investigated the resistivity $\rho$ and specific heat (by ac
calorimetry) of \ce\ in a hydrostatic helium pressure medium,
along with further resistivity measurements on a variety of
samples in a quasi-hydrostatic steatite medium. By providing some
detailed results we can perhaps provide some clues that might be
used to recognize VF mediated superconductivity elsewhere.

\begin{figure}[!ht]
\begin{center}
\includegraphics[width=0.45\textwidth]{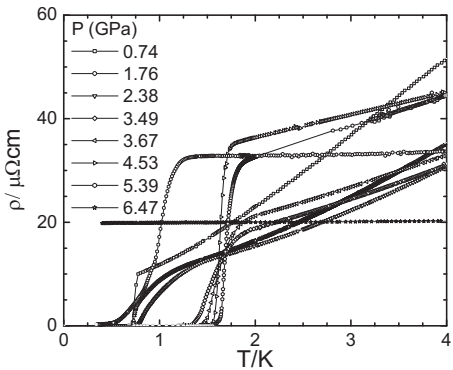}
\end{center}
\caption{Resistive superconducting transitions in \ce, in a helium
pressure transmitting medium. Note the large and pressure
dependent transition widths.} \label{fig1}
\end{figure}
In Figure~\ref{fig1} the resistance of \ce\ below 4$\:$K and up to
6.5$\:$GPa in hydrostatic conditions is shown.  The width of the
resistive superconducting transition varies strongly with
pressure. It is sharp at the lowest pressure, where \tc\ is close
to its ambient pressure value. However, when \tc\ increases
sharply around 2--3$\:$GPa, the transition width becomes very
broad, of the same order as \tc\ itself.  Close to the maximum of
the transition temperature, the transition sharpens again, though
never becoming as sharp as at ambient pressure. It should be noted
that the maximum \tc\ does not coincide with the narrowest
transition in the high pressure region.

The very broad transition at 1.76$\:$GPa was investigated in some
detail, and it was shown that the upper part of the resistance
drop could be suppressed by increasing the measurement current,
leading to a sharp transition with a width close to the ambient
pressure value. The $R=0$ point could be suppressed more rapidly
with a magnetic field than the upper part. Along with the
observation that the specific heat jump coincided with the point
at which the resistance vanished, these indicated the presence of
filamentary superconductivity with a characteristic temperature
much higher than that of the majority of the sample.

The exact details of the superconducting transition vary
substantially between samples and with pressure conditions (see
\cite{HolmesPhD} for further examples). However these broad
resistive transitions appear to be a universal feature of \ce\ at
high pressure, regardless of the pressure conditions. Let us
recall that even for the highest $T_{c}^{\rm{onset} }$ measured in
a single crystal, at 2.4$\:$K, a tail of 1\% of the normal state
resistivity remained well below 2$\:$K, vanishing only at 1.5$\:$K
(see \cite{Vargoz98}). We can note, however, that the pressure
dependence of $T_c^{R=0}$ is remarkably well reproduced between
many different samples when normalised to its value at P=0, while
$T_{c}^{\rm{onset}}$ displays much more scatter. The status of the
superconductivity of \ce~between $T_{c}^{\rm{onset}}$ and
$T_c^{R=0}$ remains mysterious.

The recently discovered HF superconductor CeNiGe$_3$
\cite{Nakashima04} displays similarly broad transition widths,
along with a large increase in residual resistivity, and is
probably the most likely candidate for valence fluctuation
mediated superconductivity in another compound. The Ce$M$In$_5$
systems have also been shown to have a fairly large pressure
dependence of their resistive transition widths, though perhaps
not to the same extent. The relative position of the magnetic and
valence instability pressures $P_c$ and \pv\ may play an important
role in the presence or otherwise of VF mediated superconductivity
in other Ce-based HF compounds. CeCu$_2$(Ge,Si)$_2$ remains the
only case in which it is clear that $P_c \ll P_{\rm v}$.


\begin{figure}[!ht]
\begin{center}
\includegraphics[width=0.45\textwidth]{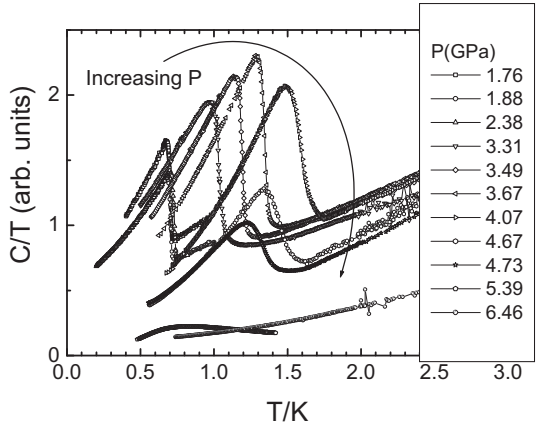}
\end{center}
\caption{Superconducting specific heat jump at different
pressures, extracted from the ac calorimetry signal and
simultaneously obtained with $\rho$, (Fig.~\ref{fig1}). Note that
it remains sharp up to 3.7$\:$GPa, but starts to broaden and
collapse even before \tc\ reaches a maximum. The amplitude of the
jump seems to be much larger closer to $P_{\rm v}\simeq4.5\:$GPa
than at low pressure near $P_c$, however there remains the
possibility that this is an experimental artefact.} \label{fig2}
\end{figure}
Figure \ref{fig2} shows the superconducting transition in \ce\ as
detected by ac calorimetry in a helium pressure medium. Details of
the experimental technique, along with some discussion its
limitations can be found in Ref.~\cite{Holmes04a}. It should be
noted that the curves shown almost certainly contain a substantial
minority contribution from the pressure medium and diamond anvils,
so precise quantitative comparisons are difficult. However, the
shape of the specific heat jump is a legitimate topic for
discussion.

As noted above, the beginning of the $C_p$ transition corresponds
precisely to the temperature at which the resistivity reaches
zero.  One should notice that the jump starts off relatively small
and sharp, then grows much bigger as $T_c$ increases, while
remaining sharp. Before $T_c$ reaches a maximum, the specific heat
peak starts to broaden and then collapses in amplitude as the
pressure is increased further and $T_c$ is driven to zero.  On
reducing the pressure, the specific heat jump reversibly regains
its shape, so we can rule out a disintegration of the sample.
Again while we emphasise that the results shown here are only
semi-quantitative, the apparent large increase in the magnitude of
$\Delta C_p /\gamma T$ around 3$\:$GPa from is sufficiently
reminiscent of the huge specific heat jump seen in CeCoIn$_5$ to
merit further investigation, if technical progress permits it.


\begin{figure}[!ht]
\begin{center}
\includegraphics[width=0.45\textwidth]{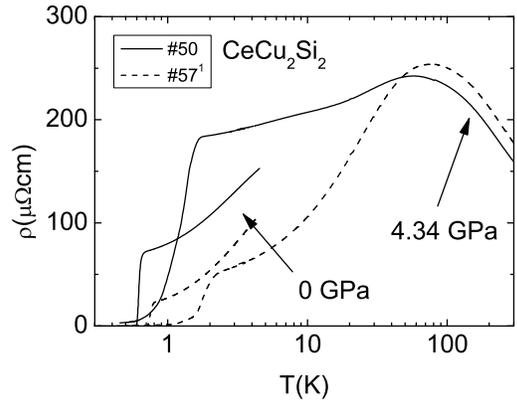}
\end{center}
\caption{Resistivity at ambient pressure and close to \pv\ of two
polycrystalline \ce\ samples prepared by Ishikawa. Note the large
increase in residual resistivity under pressure combined with a
nearly complete resistive transition.} \label{fig3}
\end{figure}

One of the most striking features of \ce\ is the huge enhancement
of the residual resistivity $\rho_0$ under pressure, which was
explained theoretically by Miyake and Maebashi \cite{Miyake02}.
Even more interesting is the coexistence of superconductivity with
such enhanced impurity scattering, a situation highly unlikely in
a spin fluctuation mediation scenario. The normal state
resistivity at \tc\ can reach of the order of the Ioffe-Regel
limit (which turns out to be about 100$\:\mu\Omega $cm at ambient
pressure).

Figure \ref{fig3} shows resistivities of two polycrystalline
samples whose compositions vary by tiny amounts within the
homogeneity range \cite{Holmesunpublished,Ishikawa03}. The
superconducting transitions are shown both at ambient pressure,
and at 4.34$\:$GPa, close to the maximum of $\rho_0$. The sample
labelled \#50 has a slightly lower \tc\ at ambient pressure than
sample \#57, along with a residual resistivity of the order of
70$\:\mu\Omega$cm if the normal state resistivity is extrapolated
to $T=0$.

At 4.34$\:$GPa, the normal state resistivity at \tc, has more than
doubled in sample \#50, and increased substantially in sample
\#57\footnote{The resistivity of \#57 under pressure has been
slightly corrected for experimental difficulties (a small constant
contact resistance was subtracted) -- see \cite{Holmesunpublished}
for further explanation.}, and the resistive transition is nearly
(though not totally) complete in both. Similar behaviour has been
seen in single crystalline \ce\ samples \cite{Jaccard98}.
Comparing the temperature dependence of the normal state, it seems
that the enhancement of the impurity scattering is reduced as
temperature increases, and the two samples show very similar
behaviour approaching room temperature.


The results discussed in the previous section indicated that there
exists another temperature dependent contribution to the
resistivity in addition to the usual electron-electron, phonon,
and magnetic scattering terms. This is the VF enhanced impurity
scattering.

According to the theoretical prediction \cite{Miyake02}, the
residual resistivity $\rho_{0}$ at $T=0$ is given by
\begin{equation} \label{eq:rho0}
\rho_{0}=Bn_{\rm imp}|u(0)|^{2}\ln\biggl|\biggl( -{\partial n_{
f}\over\partial\epsilon_{ f}}\biggr)_{\mu}/N_{\rm F}\biggr|
+\rho_{0}^{\rm unit},
\end{equation}  where the coefficient $B$ is determined by the host
metal band structure, $n_{\rm imp}$ is the concentration of
impurities with moderate scattering potential $u(q)$ coming from
disorder other than Ce ions, $N_{\rm F}$ is the density of states
of quasiparticles around the Fermi level, and the last term
represents the residual resistivity due to unitary scattering
mainly arising from any deficit or defect of the Ce ions (likely
to be small in the samples reported). The term $\left(-{\partial
n_{f}/\partial\epsilon_{f}}\right)_{\mu}$ is a measure of the rate
of valence change with pressure (which acts through the $f$-level
$\epsilon_{f}$, and is maximum at the valence instability pressure
\pv).

The temperature dependence of the enhanced impurity scattering has
not so far been predicted, and progress on this front would be
very useful, in order to better analyse any non-Fermi liquid
exponents in the quasiparticle scattering. According to
Fig.~\ref{fig3}, there should be a term with negative temperature
dependence, reminiscent of the Kondo impurity term, and indeed
such behaviour has been attributed to scattering from remaining
Ce$^{3+}$ `impurities' in the intermediate valence regime
\cite{Aliev82}.

\begin{figure}[!t]
\begin{center}
\includegraphics[width=0.45\textwidth]{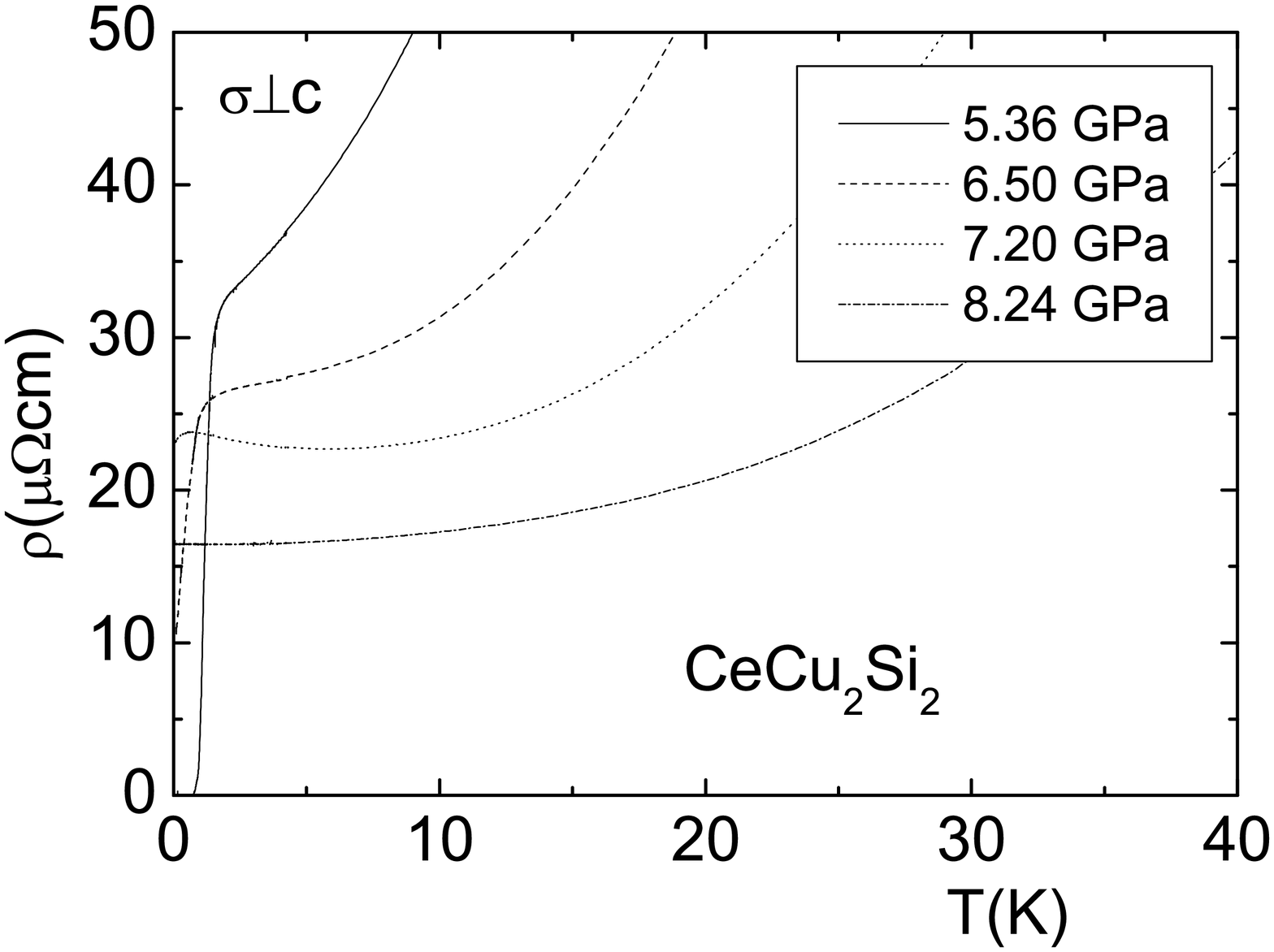}
\end{center}
\caption{The temperature dependence of the normal state
resistivity at low temperature and high pressure has two
contributions, with positive and negative slopes -- $AT^n$
quasiparticle scattering, and a temperature dependent impurity
term. Only at 7.20$\:$GPa is the negative slope of the latter
visible in this sample.} \label{fig4}
\end{figure}
Figure \ref{fig4} illustrates how a negative slope of $\rho(T)$
can occur due to the competition between the quasiparticle and
impurity scattering at high pressure, in another sample of \ce. At
the pressures shown, superconductivity is being suppressed, and
the $A$ coefficient of the quasiparticle scattering $AT^2$ is
rapidly falling, indicating the transition from a heavy fermion to
valence fluctuating regime. Only at 7.20$\:$GPa is there a clear
minimum in the resistivity, around 6$\:$K. At lower pressure, the
$A$ coefficient was large enough to mask the impurity
contribution, while at still higher pressure, the enhancement of
the residual resistivity is reduced as the system moves away from
the valence instability. One very puzzling observation in some
samples was a further change in the sign of the slope as
$T\rightarrow 0$, giving a maximum around 4$\:$K, which persists
in a magnetic field large enough to suppress any trace of
superconductivity \cite{VargozPhD}.

In summary, we have shown some unexplained features of the high
pressure superconducting and normal state of \ce\ around the
valence instability, with the aim that these might be better
explained theoretically, and also as possible signposts for
critical valence fluctuations in other compounds.

\section*{Acknowledgements}
Many thanks to K. Miyake for theoretical support, and M. Ishikawa,
C. Geibel and H. Jeevan for sample preparation. A.H. thanks the
Japan Society for the Promotion of Science fellowship programme.

%
%
%
%
%
%
%
%


\begin{thebibliography}{10}
\expandafter\ifx\csname url\endcsname\relax
  \def\url#1{\texttt{#1}}\fi
\expandafter\ifx\csname
urlprefix\endcsname\relax\def\urlprefix{URL }\fi

\bibitem{Mathur98}
N.~D. Mathur, et al., Nature \textbf{394} (1998) 39.

\bibitem{Flouquet05}
J.~Flouquet, cond-mat/0501602.

\bibitem{Yuan03b}
H.~Yuan, et al., Science \textbf{302} (2003)  2104.

\bibitem{Onishi00}
Y.~Onishi and K.~Miyake, J. Phys. Soc. Jpn. \textbf{69} (2000)
3955--3964.

\bibitem{Holmes04a}
A.~T. Holmes, et al., Phys. Rev. B \textbf{69} (2004) 024508.

\bibitem{Holmesunpublished}
A.~T. Holmes, et al., cond-mat/0505613.

\bibitem{Demuer02}
A.~Demuer, et al., J. Phys: Condens. Matter \textbf{14} (2002)
  L529.

\bibitem{bauer:147005}
E.~D. Bauer, et al., Phys. Rev. Lett. \textbf{93} (2004) 147005.

\bibitem{HolmesPhD}
A.~T. Holmes, Ph.D. thesis, University of Geneva (2004).
\newline\urlprefix\url{http://www.unige.ch/\\cyberdocuments/theses2004/HolmesAT/\\%
these.pdf}

\bibitem{Vargoz98}
E.~Vargoz, et al., Solid State Commun. \textbf{106} (1998) 631.

\bibitem{Nakashima04}
M.~Nakashima, et al., J. Phys: Condens. Matter \textbf{16} (2004)
L255.

\bibitem{Miyake02}
K.~Miyake and H.~Maebashi, J. Phys. Soc. Jpn.
  \textbf{71} (2002) 1007.

\bibitem{Ishikawa03}
M.~Ishikawa, et al., Phys. Rev. B \textbf{68}
  (2003) 024522.

\bibitem{Jaccard98}
D.~Jaccard, et al., Rev. High Pressure Sci. Techno. \textbf{7}
(1998) 412.

\bibitem{Aliev82}
F.~G. Aliev, et al., Sov. Phys. Solid State \textbf{24} (1982)
1786.

\bibitem{VargozPhD}
E.~Vargoz, Ph.D.
  thesis, University of Geneva (1998).

\end{thebibliography}
\end{document}